\documentclass{aa}    
\def\ut#1{\mathop{\vtop{\ialign{##\crcr
     $\hfil\displaystyle{#1}\hfil$\crcr\noalign
     {\kern1pt\nointerlineskip}\hbox{$\hfil\sim\hfil$}\crcr
     \noalign{\kern1pt}}}}}

\def\undersymbol#1#2{\mathop{\vtop{\ialign{##\crcr
     $\hfil\displaystyle{#2}\hfil$\crcr\noalign
     {\kern1pt\nointerlineskip}\hbox{$\hfil#1\hfil$}\crcr
     \noalign{\kern1pt}}}}}
\def\arcsec{^{\prime\prime}}

\begin{document}
\thesaurus{02.05.1 ; 02.07.1 ; 10.03.1}
\title{Astrophysical constraints on a possible neutrino ball at the Galactic Center}
\author{
    F. De Paolis
\inst{1},
    G. Ingrosso
\inst{1},
  A. A. Nucita
  \inst{1},
   D. Orlando
\inst{1},
    S. Capozziello
\inst{2} \and  G. Iovane \inst{2}
     }
\institute{Dipartimento di Fisica, Universit\`a di Lecce, and
INFN, Sezione di Lecce, Via Arnesano, CP 193, I-73100 Lecce, Italy
\and Dipartimento di Fisica ``E.R. Caianiello", Universit\`a di
Salerno, and INFN, Sezione di Salerno, Via S. Allende, I-84081
Baronissi (Sa),  Italy } \offprints{G. Ingrosso}
\date{Received date; accepted date}
\authorrunning{F. De Paolis et al.}
\titlerunning{Neutrino ball at the Galactic Center} \maketitle
\begin{abstract}

The nature of the massive object at the  Galactic Center (Sgr
A$^{*}$) is still unclear even if various observational campaigns
led many authors to believe that our Galaxy hosts a super-massive
black hole with mass $M\simeq 2.6\times 10^6$ M$_{\odot}$.
However, the black hole hypothesis, which theoretically implies a
luminosity $\simeq 10^{41}$ erg s$^{-1}$, runs into problems if
one takes into account that the observed luminosity, from radio
to $\gamma$-ray wavelengths, is below $10^{37}$ erg s$^{-1}$. In
order to solve this blackness problem, alternative models have
been recently proposed. In particular, it has been suggested that
the Galactic Center hosts a ball made up by non-baryonic matter
({\it e.g.} massive neutrinos and anti-neutrinos) in which the
degeneracy pressure of fermions balances their self-gravity.
Requiring to be consistent with all the available observations
towards the Galactic Center allows us to put severe astrophysical
constraints on the neutrino ball parameters. The presence of such
an object in the Galactic Center may be excluded if the
constituent neutrino mass $m_{\nu}$ is $\ut> 24$ keV, while if
$m_{\nu}\ut< 24$ keV observations can not give a definite answer.

\keywords{Elementary Particles, Gravitation, Galaxy Center}
\end{abstract}

\section{Introduction}
There is much evidence for the presence of super-massive black
holes (SBHs) with masses in the range $10^6 - 10^9~M_{\odot}$ in
QSOs, AGNs and centers of galaxies. In our Galaxy, the discovery
of the unusual radio source SgrA$^*$ and the detailed information
coming from star dynamics led many authors to believe that also
our Galaxy hosts a SBH with mass $(2.6 \pm 0.4)\times 10^6 ~
M_{\odot}$ (Genzel, Thatte, Krabbe and Tacconi-Garman
\cite{gtkt}). Since this SBH should accrete the surrounding gas
at the rate $\dot{M}\simeq 10^{-6}~M_{\odot}$ yr$^{-1}$ (Ghez,
Klein, Morris and Becklin \cite{gkmb}), the usual radiative
efficiency $\epsilon \simeq 0.1$ would imply a luminosity $L\geq
10^{41}$ erg s$^{-1}$. However, the observed luminosity from
radio to $\gamma$-ray wavelengths is below $10^{37}$ erg s$^{-1}$,
thereby implying that the SBH hypothesis runs into problems
(Goldwurm, Cordier, Paul et al. \cite{gold}). This is the so
called ``blackness problem'' or the ``black hole on starvation''.
Several alternative models have been proposed to solve the issue.
For example, Narayan, Mahadevan, Grindlay  et al. (\cite{nmg})
proposed the advection dominated accretion flow model (ADAF),
according to which most of the dissipated energy is channeled
into protons that cannot radiate efficiently. However, as has
been recently noted, the polarized radiation from SgrA$^*$
requires a nonthermal electron distribution for the emitting
plasma and this seems to imply that the ADAF model is ruled out
(Agol \cite{agol}). In principle, the direct observation of a
mass density profile $\rho (r)\propto r^{-7/4}$ (Binney and
Tremaine \cite{bt}) very close to SgrA$^*$ should allow to
confirm the presence of a compact object like a SBH at the
Galactic Center. However, the present observational techniques do
not permit to distinguish stars at distances $r \ut< 10^{-2}$ pc
from the Galactic Center, so that the SBH hypothesis at the
Galaxy Center is far from being conclusive\footnote{For different
methods to investigate the nature of the Galactic Center see {\it
e.g.} De Paolis, Gurzadyan and Ingrosso (\cite{dgi}), De Paolis,
Ingrosso and Nucita (\cite{din}), and references therein.}.

Recently, Torres, Capozziello and Lambiase (\cite{tcl}) have
investigated the hypothesis that the Galactic Center could consist
of a super-massive boson star. They analyzed  stability
configurations and dynamics  giving the prospects for the
observational detection  of such an object, using the new
generation of X-ray, radio interferometry satellites and, in
general, tools capable of detecting strong gravitational lensing
effects. The conclusions were that the SBH hypothesis is, again,
far from being definitive while the ``signature" of boson stars
could be (or not) soon available in the case of very massive
bosons (see {\it e.g.} Capozziello, Lambiase and Torres
\cite{clt}).

Another alternative model to the SBH scenario at the Galactic
Center has been  proposed by some authors (Tsiklauri and Viollier
\cite{viollier1}, Tsiklauri and Viollier \cite{viollier2},
Capozziello and Iovane \cite{ci}). According to this scenario,
neutrinos and anti-neutrinos could gravitationally interact
forming super-massive neutrino balls in which the degeneracy
pressure of fermions balances the self-gravity of the system.

Choosing neutrinos in a particular mass range implies the
formation of super-massive fully degenerate objects with mass
$\simeq 10^6$ M$_{\odot}$. In particular, Capozziello and Iovane
\cite{ci} proposed to investigate the eventual neutrino ball at
Galactic Center by gravitational lensing, since the neutrino ball
acts as a transparent medium.

On the other hand, the existence of a neutrino ball at the
Galactic Center avoids to invoke the presence of a SBH and, under
certain circumstances, should be able to justify the low
luminosity (from radio to $\gamma$-rays) observed towards
SgrA$^*$.

However, if a neutrino ball really exists at the center of the
Galaxy, this possibility  must be consistent both with the
theoretical mass limit of a stable configuration of fermions and
with the currently available observational data, i.e. {\it i)}
the star dynamics within about one pc from SgrA$^*$, {\it ii)}
the low source luminosity. In addition, the interaction among
neutrinos and anti-neutrinos within the ball, or the decay of
neutrinos into neutrinos of different flavors, may also produce
characteristic signatures revealing the object at the Galactic
Center.

The aim of the present paper is to derive astrophysical
constraints on the parameters of the neutrino ball with
particular attention to the neutrino mass $m_{\nu}$. To this
purpose, we relax the assumption of fully degenerate neutrino
configurations adopted until now in the literature. Accordingly,
we adopt a formalism based on the distribution function in
phase-space allowing to obtain more general fermion
configurations with a degeneracy degree depending on the radial
coordinate within the ball. In this formalism either classical
configurations (in which particles obey the Maxwellian statistics)
and fully degenerate systems are naturally included. In this way,
considering in addition the astrophysical constraints {\it i)}
and {\it ii)} in the paragraph above, the allowed neutrino mass
range turns out to be 11 keV $\ut<m_{\nu}\ut<$ 24 keV.

We note that assuming neutrinos in this mass range, it is also
possible to build up neutrino ball models with total mass up to
$10^9-10^{10}$ M$_{\odot}$ and radius $\simeq 10^{-3}-10^{-2}$
pc. These objects might influence the accretion process of
super-massive black holes in the AGN cores or completely mimic
the central black holes, acting as the engine of AGNs. This
possibility has been explored in some details by Tsiklauri and
Viollier (\cite{tsiklauri}), but considering only fully
degenerate self-gravitating configurations.

A further problem to be addressed is the cosmological implications
of the existence of such heavy neutrinos. This problem has been
discussed by several authors (e.g. Kolb and Turner
\cite{kolbeturner}, Viollier \cite{viollierCosmo}, Lindebaum,
Tupper and Viollier \cite{ltv}, Dolgov and Hansen \cite{dh}) to
whom we refer for further details. Here we mention that an active
neutrino ($\nu_e$, $\nu_{\mu}$ or $\nu_{\tau}$) of mass of a few
keV  is the warm dark matter candidate preferred by many authors,
on the basis of N-body simulations of large scale structure
formation (e.g. Colin, Avila-Reese and Valenzuela \cite{cav}).
Indeed, Big Bang nucleosynthesis can only exclude active neutrino
masses bigger than about $300$ keV (Dolgov, Hansen and Semikoz
\cite{dhs}). However, in the framework of the standard cosmology,
active neutrinos with mass in the range  11 keV $\ut<m_{\nu}\ut<$
24 keV, overclose the universe by a factor of about $100$ (Kolb
and Turner \cite{kolbeturner}). Consequently, if heavy neutrinos
formed and were in equilibrium in the early universe, they have
to rapidly decay in order to not overclose it. Therefore,
standard cosmology strongly constrains the presence of heavy
neutrinos nowadays. Many authors have discussed this issue in the
framework of more exotic cosmological scenarios. Indeed, it has
been shown that the cosmological bound on neutrino mass can be
bypassed at least in three ways: by {\it i)} avoiding to
thermalize massive neutrinos with a reheat temperature (Giudice,
Kolbe, Riotto et al. \cite{gkrs}), {\it ii)} decay of neutrinos,
{\it iii)} annihilation of neutrinos and anti-neutrinos
(Bili\'{c}, Munyaneza and Viollier \cite{bmv} and references
therein). Another possibility has been explored by Dolgov and
Hansen (\cite{dh}) who have shown that in the framework of a
slightly extended standard model of elementary particles, right
handed (or sterile) neutrinos of mass $\ut < 20$ keV, are not in
contrast with cosmological constraints (see also Shi and Fuller
\cite{shifuller}). However, this issue cannot be considered
firmly established since this model is constrained by
astrophysical bound on $\nu_s\rightarrow \nu \gamma$ (see e.g.
Dress and Wright \cite{dw}). Alternatively, in contrast with the
standard cosmology, some authors (Viollier \cite{viollierCosmo},
Bili\'{c}, Lindebaum, Tupper and Viollier \cite{bltv}, Lindebaum,
Tupper and Viollier \cite{ltv}) have proposed a scenario
according to which the universe has become heavy neutrino matter
dominated $22$ days after the Big Bang at temperature $\sim$ 1
keV. From that time on, the evolution of the universe differs
substantially from the standard cosmology results since the
universe will undergo a gravitational phase transition leading to
super-massive neutrino systems with masses close to the
Oppenheimer-Volkoff limit $\simeq 3\times 10^9$ M$_{\odot}$
(Lindebaum, Tupper and Viollier \cite{ltv} and references
therein). At this stage, annihilation of heavy neutrinos into
non-standard light bosons may take place, thereby reducing the
neutrino number density inside neutrino systems.

However, well aware that in the framework of standard cosmology
the heavy neutrino hypothesis meets with difficulties, in the
present paper we focus on a set of independent astrophysical
constraints on the neutrino mass that can be derived from the
observational data towards the galactic center. Obviously,
further theoretical analysis is necessary in order to clarify if
heavy neutrinos may really exist and cluster in the galactic
centers. Anyway, the next generation of $X$-ray and $\gamma$-ray
satellites, with improved sensitivity and angular resolution, will
allow to definitely confirm or exclude the presence of a massive
neutrino ball at the galactic center.

The outline of the paper is the following: in Section 2 we
describe the adopted neutrino ball model, then we consider in
Section 3 a set of astrophysical constraints which can be put on
the neutrino ball parameters. In Section 4, we investigate the
observational signatures from this exotic object at the Galactic
Center. Our main conclusions are summarized in Section 5.

\section{The neutrino ball model}

The gravitational equilibrium of a fully degenerate system of
fermions is well known since Chandrasekhar (\cite{chandrasekhar})
who showed that equilibrium configurations do exist if the total
number of particles is less than the critical value $N_{crit}
\simeq (m_{Planck}/m_{\nu})^3$. Here $m_{Planck}\simeq (\hbar
c/G)^{1/2}$ is the Planck mass. However, it has been stressed by
several authors that in the standard cosmology a degeneracy value
near zero (corresponding to the semi-degenerate case) should be
expected for neutrinos produced in the early universe (see e.g.
Dolgov and Zel'dovich \cite{dz}). When the gravitational
configurations of semi-degenerate systems of fermions are
calculated, a spatial divergence appears and the solutions are
not finite in masses and radii  as in the case of isothermal
systems obeying the classical statistics (Gao and Ruffini
\cite{gr}). A solution to this problem has been proposed by
Ruffini and Stella (\cite{rs}) on the basis of the early work of
King (\cite{king}).
 These works introduce a distribution function modified with an energy
cutoff in phase space and allow to obtain self-gravitating systems
limited in extension since the velocity of the particles at any
point of the system has to be lower than the escape velocity. In
the case of spherical symmetry and within the non-relativistic
approximation (see below), this escape velocity is given in terms
of the gravitational potential $V(r)$ by
\begin{equation}
v_e^2(r) = - 2 V(r)~,
\end{equation}
where $V(r)$ is fixed to be zero at the boundary $R$ of the
system.

Massive neutrinos are considered to be collisionless and are
described in the momentum space by the distribution function
(Ruffini and Stella \cite{rs})
\begin{equation}
\left\{
\begin{array}{l}
dn(r)=\frac{g}{h^3}~\frac{1-e^{({\epsilon-\epsilon_c(r)})/kT}}
{e^{({\epsilon - \mu(r)}) /kT} + 1} d^3 p(\epsilon)~~~{\rm
for}~~~ \epsilon\leq \epsilon_c(r) \\ \\ \nonumber dn(r)=0
~~~~~~~~~~~~~~~~~~~~~~~~~~~~~~~~{\rm for}~~~ \epsilon
>\epsilon_c(r)\ ,
\end{array}
\right. \label{df}
\end{equation}
where $\epsilon_c(r)$ is the energy cutoff for neutrinos,
$g=2s_{\nu}+1$ the spin multiplicity of quantum states, $\mu(r)$
the chemical potential and $T$ the fermionic thermodynamic
temperature assumed to be constant. This distribution function
generalizes to semi-degenerate systems of fermions the
distribution function introduced by King (\cite{king}) for
systems of classical particles (for which $\mu(r) \to -\infty$).
At the same time it includes the usual Fermi-Dirac statistics,
which is recovered from equation (\ref{df}) in the limit $\mu(r)
\to + \infty$ and $\epsilon_c(r) \to + \infty$.

In view of the application of the distribution function in
equation (\ref{df}) to the neutrino ball at the Galactic Center,
we restrict our attention to the non-relativistic regime because
of the densities involved in this structure. With the
approximation $\epsilon = m_{\nu} v^2/2$ and $kT << m_{\nu}c^2$,
after integration on $d^3p$, one gets the mass density $\rho(r)$
and the pressure $p(r)$ as a function of the radial coordinate $r$
\begin{equation}
\rho(r) = {2 \pi g~m_{\nu}^4  \over h^3 j^3}~\int_0^{W(r)} {
1-e^{x-W(r)} \over e^{x-\theta(r)}+1 }~x^{1/2}~dx~.
\label{density}
\end{equation}
\begin{equation}
p(r) = \frac{2}{3} \frac{\pi g~m_{\nu}^4} {h^3 j^5} ~\int_0^{W(r)}
{ 1-e^{x-W(r)} \over e^{x-\theta(r)}+1 }~x^{3/2}~dx~,
\label{pressure}
\end{equation}
where
\begin{equation}
x = j^2 v^2~,~~~~~~j^2 = \frac{m_{\nu}}{2kT}
\end{equation}
and the quantities $W(r)$ and $\theta(r)$ are the energy-cutoff
and degeneracy parameter defined by
\begin{equation}
W(r) = \frac{\epsilon_c(r)}{kT}~~~~{\rm and}~~~
\theta(r)=\frac{\mu(r)}{kT}~.
\end{equation}
We note that in the fully degenerate limit, being $\theta(r)\gg
1$ and $W(r)\gg 1$, equations (\ref{density}) and
(\ref{pressure}) give the usual density and pressure of a
degenerate Fermi gas (Chandrasekhar \cite{chandrasekhar})
\begin{equation}
[\rho(r)]_{deg}={4\pi g m_{\nu}^4\over 3h^3}\left({2\mu (r)\over
m_{\nu}}\right) ^{3/2}~, \label{densitydegenere}
\end{equation}
\begin{equation}
[p(r)]_{deg}={4\pi g m_{\nu}^4 \over 15 h^3} \left({2\mu (r)\over
m_{\nu}}\right)^{5/2}~. \label{pressuredegenere}
\end{equation}
Density and pressure in equations (\ref{density}) and
(\ref{pressure}) depend on four parameters: the mass of the
particle $m_{\nu}$, the temperature parameter $j$, the energy
cutoff parameter $W(r)$ and the degeneracy factor $\theta(r)$.
However, the assumed condition of thermal equilibrium in the
presence of a gravitational field for the neutrino ball implies
the following relation between $\theta(r)$ and $W(r)$ (Ingrosso,
Merafina, Ruffini and Strafella \cite{imrs})
\begin{equation}
\theta(r) = W(r) + \theta_R~,~ \label{bordo}
\end{equation}
where $\theta_R\leq 0$ is the degeneracy evaluated at the system's
surface.

Assuming spherical symmetry, within the non-relativistic limit,
the equations governing the gravitational equilibrium of the
self-gravitating systems are
\begin{equation}
\frac{dp}{dr} = - \frac{G M(r) \rho }{ r^2}~,~~~~~ \frac{dM}{dr} =
4 \pi r^2 \rho~, \label{equilibrio}
\end{equation}
which can be recasted in the form
\begin{equation}
\frac{dW}{dr} = -2 j^2 \frac{GM(r)}{r^2}~,
\end{equation}
from which it follows that
\begin{equation}
\frac{d^2W}{d \xi^2} + \frac{2}{\xi} \frac{dW}{d\xi} = -8 \pi G
j^2 r_0^2 \rho~, \label{poisson}
\end{equation}
where $\xi = r/r_0$ is a dimensionless radius expressed in terms
of the ``core radius''
\begin{equation}
r_0^2 = \frac{9} {8 \pi G \rho(0)j^2}~. \label{formula}
\end{equation}
We note that the particle velocity dispersion obeys the relation
\begin{equation}
<v^2(r)> =\frac{3p(r)}{\rho (r)j^2}~, \label{formula2}
\end{equation}
which in the classical limit reduces to the Boltzmann relation
$<v^2(r)>=3/(2j^2)$, while in the fully degenerate case reads
$<v^2(r)>=3W(r)/(5 j^2)$.

Equation (\ref{poisson}) has to be integrated with the boundary
conditions $W(0)=W_0$ and $W^{\prime}(0)=0$ from the center of
the configuration to the surface at which $W(R)=0$. Clearly,
equation (\ref{poisson}), through the definition of the density
$\rho(r)$, also depends on the degeneracy parameter at the surface
of the configuration $\theta_R$, while the dependence on the
parameters $m_{\nu}$ and $j$ disappears by the definition of
$r_0$. The radius $R$ and the total mass $M$ of the system are
given by
\begin{equation}
R = r_0 \xi_R~~~~~~{\rm and}~~~~~~~ M= \frac{r_0}{2 G j^2} \left(
-\xi^2 \frac{dW}{d\xi}\right)_{\xi = \xi_R}~, \label{radius&mass}
\end{equation}
where $\xi_R$ is the value of the radial coordinate at which
$W(\xi_R) = 0$ and $\rho(\xi_R) =0$.

\begin{figure}
\begin{center}
\vspace{9.cm} \includegraphics{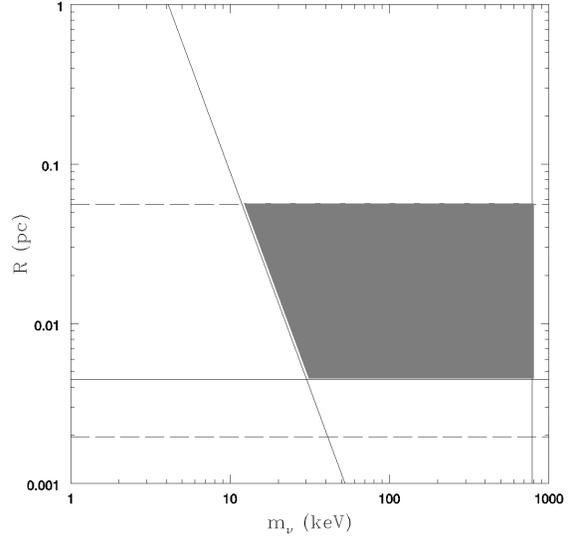} \caption{The neutrino
ball radius $R$ is reported as a function of the constituting
neutrino mass $m_{\nu}$. The neutrino ball mass has been assumed
to be $M=(2.6\pm0.2)\times 10^6$ M$_{\odot}$. The oblique dashed
line corresponds to the full degenerate models. The vertical
continuous line represents the Chandrasekhar limit. The two
horizontal dashed lines bound the models satisfying the observed
luminosity towards SgrA$^*$. The models above the horizontal
solid line have $T_{ev}\geq T_H$. Details are throughout the text.
 } \label{rcontrom}
\end{center}
\end{figure}

A detailed analysis of the numerical solutions of equation
(\ref{poisson}) has been performed by Ruffini and Stella
(\cite{rs}) and  Ingrosso, Merafina, Ruffini and Strafella
(\cite{imrs}), showing that simple scaling relations between
$m_{\nu}$, $M$ and $R$ may be found both in the classical and in
the degenerate cases. In fact, in the classical limit $\theta_R
\to - \infty$, one gets
\begin{equation}
m_{\nu} \propto R^{-3/8} e^{-\theta _R/4} M^{-1/8}~.
\label{msemidegenerate}
\end{equation}
In the degenerate limit the dependence of $m_{\nu}$ on the
parameter $\theta_R$ disappears so that the scale law is
\begin{equation}
m_{\nu} \propto R^{-3/8} M^{-1/8}~, \label{mdegenerate}
\end{equation}
corresponding to the well known Chandrasekhar's
(\cite{chandrasekhar}) scaling laws. The typical relationship
between the ball radius and the constituting neutrino mass is
shown in Figure \ref{rcontrom} and it will be discussed in the
following Section.

\section{Astrophysical constraints}

The aim of this Section is to determine the constraints that
current astrophysical observations can put on the physical
parameters of a possible neutrino ball at the Galactic Center,
described by the model outlined in Section 2. These constraints
are the consequence both of the available astronomical
observations, like star dynamics within $1$ pc from the Galactic
Center and the SgrA$^*$ luminosity observed in all wavelenghts,
and of the theoretical mass limit for stable configurations of
fermions.

First of all, it is well known that the two dimensional positions
and velocities measured for stars in the inner
$6\arcsec~\times~6\arcsec$ (0.23 $\times$ 0.23 pc) provide
excellent constraints on the matter distribution at the galactic
center. With the assumption that stars are gravitationally bound
by the central gravitational potential, Bahcall and Tremaine
(\cite{bat}) proposed to use a projected mass estimator which, for
the case of star isotropic orbits, is given by
\begin{equation}
M = \frac{16}{\pi G}<v^2(b) b> \label{estimator}
\end{equation}
where $v(b)$ and $b$ represent the star projected velocity and
distance from SgrA$^*$, respectively. For the Galactic Center data
the estimator above gives, for the amount of matter contained
within $0.015$ pc, the value $M=(2.6\pm0.2)\times 10^6$
M$_{\odot}$ (Ghez, Klein, Morris and Becklin \cite{gkmb}).

As far as the SgrA$^*$ luminosity is concerned, Melia
(\cite{melia}) showed that observations of stellar winds and gas
flows near SgrA$^*$, coupled with the above estimate of the SBH
mass, imply a minimum mass accretion rate $\dot{M}\simeq 10^{-6}$
M$_{\odot}$ yr$^{-1}$ (Genzel, Hollenbach and Townes \cite{ght}).
This estimate, for a standard thin accretion disk around the SBH
would imply a total luminosity $L\geq 10^{41}$ erg s$^{-1}$ with
a peak emission in the $X$-ray band. On the contrary, the total
luminosity observed from radio to $\gamma$-ray wavelengths is
below $10^{37}$ erg s$^{-1}$, peaked in the near infrared region
corresponding to a photon energy $E_{\gamma}\simeq 5\times
10^{-5}$ keV (Narayan, Mahadevan, Grindlay et al. \cite{nmg}).

In the neutrino ball scenario, assuming that the star dynamics
around SgrA$^*$ is accounted for by the galactic neutrino ball
gravitational potential, we need that the mass enclosed within
about $10^{-2}$ pc is $M\simeq 2.6\times 10^6$ M$_{\odot}$. This
condition, as it is evident from equations (\ref{msemidegenerate})
and (\ref{mdegenerate}), allows us to determine the neutrino mass
as a function of the ball radius in the case of degenerate
systems, and of both the ball radius and degeneracy parameter
$\theta_R$ in the case of semi-degenerate ones. This is shown in
Figure \ref{rcontrom} where the oblique dashed line corresponds
to the fully degenerate systems. The models on the right part
with respect to this line are models with decreasing values of
the degeneracy parameter $\theta_R$ and increasing values of the
neutrino mass $m_{\nu}$.

An upper limit to the constituting neutrino mass $m_{\nu}$ does
anyway exist as a consequence of General Relativity. In fact, it
is well known (see e.g. Shapiro and Teukolsky \cite{st}) that the
balance between the gravitational force and the degeneracy
pressure leads to stable configurations of fermions until the
number of particles composing the system does not exceed the
critical value given by
\begin{equation}
(N_{\nu})_{max}\simeq \left(\frac{\hbar
c}{G}\right)^{3/2}m_{\nu}^{-3}~. \label{maxnumber}
\end{equation}
In this way, the maximum neutrino mass allowed by the General
Relativity can be estimated as a function of the total mass by
\begin{equation}
(m_{\nu})_{max}\simeq 787\left(\frac{2.6 \times 10^6~{\rm
M_{\odot}}}{M}\right)^{\frac{1}{2}}~{\rm keV}~, \label{grlimit}
\end{equation}
which, for the assumed neutrino ball mass, entails
$m_{\nu,max}\simeq 787$ keV. This limit corresponds to the
vertical solid line in Figure \ref{rcontrom}.

In addition to the above dynamical constraint, also the SgrA$^*$
luminosity has to be consistent with the ball parameters $M$, $R$
and $m_{\nu}$. We assume that the luminosity observed from the
Galactic Center is the result of the accretion of the surrounding
gas on the neutrino ball. In this case, for a spherical inflow,
the energy of the emitted thermalized photons is given by
(Shapiro and Teukolsky \cite{st})
\begin{equation}
E_{\gamma}\simeq 10^{-5}~R_{16}^{-1/2}L_{37}^{1/4}~~~{\rm keV}~,
\label{photonenergy1}
\end{equation}
where $R_{16}$ and $L_{37}$ represent the neutrino ball radius and
the total luminosity in units of $10^{16}$ cm and $10^{37}$ erg
s$^{-1}$, respectively.

Therefore, for an observed SgrA$^*$ luminosity $L \simeq 10^{37}$
erg s$^{-1}$, and for a photon energy $E_{\gamma} \simeq (5 \pm 1)
\times 10^{-5}$ keV, the previous equation entails a set of
acceptable neutrino ball models corresponding to the region
between the two horizontal dashed lines in Figure \ref{rcontrom}.
As one can see, the lower limit to the neutrino mass is
$m_{\nu}\simeq 11$ keV corresponding to the maximum acceptable
neutrino ball radius $R\simeq 5.5\times 10^{-2}$ pc. We note also
that the mass enclosed within $0.015$ pc is $2.42\times 10^6$
$M_{\odot}$, in agreement with the observational constraints.

An additional constraint on the allowed neutrino ball parameters
$R$ and $m_{\nu}$ derives by requiring that the evaporation
time-scale of the system is greater than the Hubble time
$T_H\simeq 1.4\times 10^{10}$ yrs. In fact, by considering the
interaction among neutrinos and anti-neutrinos of the ball itself
via the following reaction channels (Boehm and Vogel \cite{bv})
\begin{equation}
\begin{array}{l}
\nu_{\tau}+ \bar{\nu}_{\tau}\rightarrow \nu_{e}+\bar{\nu}_{e}\\ \\
\nu_{\tau}+ \bar{\nu}_{\tau}\rightarrow
\nu_{\mu}+\bar{\nu}_{\mu}~,
\end{array}
\label{channels1}
\end{equation}
the evaporation time-scale can be defined as
\begin{equation}
T_{ev}\simeq \frac{1}{n_{\nu}<v_{\nu}>\sigma_{\nu \nu}}~,
\end{equation}
where $\sigma_{\nu \nu} \simeq G_f^2 m_{\nu_{\tau}}^2/2 \pi \hbar
^4 g$ ($G_f$ is the Fermi coupling constant) is the process cross
section (Holdom and Malaney \cite{hm}) and the neutrino mean
velocity $<v_{\nu}>$ is given by equation (\ref{formula2}). By
requiring that $T_{ev}\geq T_H$ we obtain that the allowed models
are those above the horizontal solid line in Figure \ref{rcontrom}
corresponding to a minimum neutrino ball radius of $4.3\times
10^{-3}$ pc.

At this stage, the allowed neutrino ball parameters are
represented by the grey region in Figure \ref{rcontrom}. In the
next Section, the allowed region will be further reduced by
considering the available observations both in the $X$-ray and
$\gamma$-ray energy bands towards SgrA$^*$.

\section{Search for galactic neutrino ball signatures}

In order to study the influence of a neutrino ball at the Galaxy
Center, we investigate the observable signatures that should be
produced {\it i)} in the interaction of incoming high energy
neutrinos (or anti-neutrinos) with anti-neutrinos (or neutrinos)
composing the ball, and {\it ii)} in the interaction between
neutrinos and anti-neutrinos in the ball itself.

Let us first consider the case {\it i)}. The existence of an high
energy $\mu$ neutrino (and/or anti-neutrino) flux is a theoretical
consequence of the decay of charged pions produced in high energy
$pN$ interactions. From Cosmic Ray observation on Earth, Waxman
and Bahcall (\cite{wb}) derive an upper bound to the high energy
neutrino flux given by
\begin{equation}
\frac{dN_{\nu_{\mu}}}{dE_{\nu_{\mu}}} \simeq 2\times 10^{-8}~
\left( \frac{E_{\nu_{\mu}}} {{\rm GeV}} \right)^{-2} ~~~\frac{{\rm
neutrinos}}{{\rm cm^2~s~sr~GeV}}
\end{equation}
where $E_{\nu_{\mu}}$  represents the energy of the incident
$\mu$ neutrinos. The previous relation holds for $E_{\gamma}\geq
10^3$ GeV. In the framework of the Standard Model for weak
interactions, these particles interact with the massive
anti-neutrinos $\bar{\nu}_{\tau}$ (or neutrinos $\nu_{\tau}$)
composing the ball, giving rise to a photon flux on Earth, as a
consequence of the final neutral pion decays (Fargion, Mele and
Salis \cite{fms}). Indeed, for a $\nu_{\mu}~\bar{\nu}_{\tau}$ (and
charge conjugated $\bar{\nu}_{\mu}~\nu_{\tau}$) interaction via
the $W$ exchange in the {\it t}-channel, the cross section is
given by (Fargion, Mele and Salis \cite{fms})
\begin{equation}
\begin{array}{l}
\nonumber \sigma (s)\simeq 108.5~
\frac{A}{s}\left\{1+\frac{m^2_W}{s}\times \right.
\\
\left.~~~~~~~~~~~~~~~~~~~~~~~~~~~~~~~~\left[2-\frac{s-B}{A}
\ln\left(\frac{B+A}{B-A}\right)\right]\right\}~{\rm pb}~,
\end{array}
\end{equation}
where $m_W$ is the $W$ boson mass, $\sqrt{s}$ is the center of
mass energy which -- in the relativistic limit reads out to be
$\sqrt{s}\simeq \sqrt{2m_{\nu_{\tau}}E_{\nu}}$ -- and the two
functions $A$ and $B$ are defined, respectively, as
\begin{equation}
\left\{
\begin{array}{l}
A=\sqrt{[s-(m_{\nu_{\tau}}+m_{\nu_{\mu}})^2][s-(m_{\nu_{\tau}}-m_{\nu_{\mu}})^2]}~,\\\\
B=s+2m^2_W-m^2_{\tau}~,
\end{array}
\right. \label{parametrizzazioni}
\end{equation}
where $m_{\nu_{\tau}}$ and $m_{\nu_{\mu}}$ are the neutrino
$\nu_{\tau}$ and the neutrino $\nu_{\mu}$  masses, respectively.
From the {\it t}-channel reaction chain for final photon
production, it is easy to observe that the energy of each
incident neutrino is a multiple $\eta$ of the energy of the
produced photon ($E_{\nu_{\mu}}\simeq \eta E_{\gamma}$) and that
each step of the chain occurs with probability $P_i$ (Fargion,
Mele and Salis \cite{fms}).

\begin{figure}[htbp]
\begin{center}
\vspace{7.8cm} \includegraphics{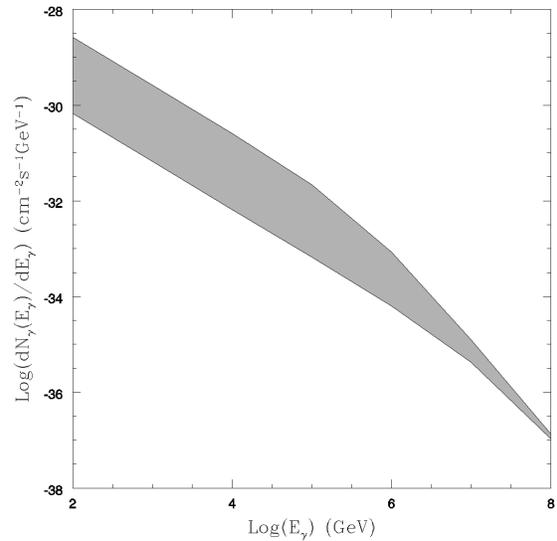} \caption{The photon flux
$dN_{\gamma}(E_{\gamma})/d E_{\gamma}$ on Earth, resulting from
the interaction of high energy cosmic neutrinos (or
anti-neutrinos) with the anti-neutrinos (or neutrinos) of the
ball, is reported as a function of the photon energy
$E_{\gamma}$. The dashed region corresponds to different photon
fluxes depending on the constituting neutrino mass in the range
$20-787$ keV. Due to the extremely low flux on Earth, this kind
of signature can not be used in order to test the presence of the
neutrino ball at the center of the Galaxy.} \label{figuraflux}
\end{center}
\end{figure}

The photon flux on Earth, obviously, depends on the neutrino ball
parameter, {\it i.e.} the radius $R$ and the neutrino number
density $n_{\nu_{\tau}} (r) \simeq\rho(r)/m_{\nu_{\tau}}$. In
this way, the photons flux at energy $E_{\gamma}$ is estimated to
be
\begin{equation}
\frac{dN_{\gamma}}{d E_{\gamma}} \simeq
\int_{0}^{R}\frac{dN_{\nu_{\mu}}}{dE_{\nu_{\mu}}}
\left(\frac{r}{D}\right)^2 e^{\sigma (s)n_{\nu_{\tau}}(r-R)}\sigma
(s)n_{\nu_{\tau}}\prod_iP_idr~, \label{highenergynunu}
\end{equation}
where $D\simeq 8.5$ Kpc is the Earth distance from the Galactic
Center. In Figure \ref{figuraflux} we show the flux
$dN_{\gamma}/d E_{\gamma} $ for different values of $E_{\gamma}$
and $m_{\nu_{\tau}} \ut> 20$ keV, assuming for the neutrino ball
mass and radius the values $M \simeq 2.6\times 10^{6}$
M$_{\odot}$ and $R = 10^{-2}$ pc, respectively. In all cases, the
$\gamma$-ray flux on Earth is too low to be measured by the
actual instrumentation leaving open the problem to discriminate
the Galactic Center dark object by using this kind of signature.

If a neutrino ball really exists at the Galactic Center, other
possible characteristic signatures come from the interaction among
neutrinos and anti-neutrinos of the ball itself via the radiative
reaction channel and by the $\tau$ neutrino decay (Boehm and
Vogel \cite{bv})
\begin{equation}
\begin{array}{l}
(a)~~\nu_{\tau}+ \bar{\nu}_{\tau}\rightarrow \gamma + \gamma\\ \\
(b)~~\nu_{\tau} \rightarrow \nu_{e}+\gamma~.
\end{array}
\label{channels}
\end{equation}
Following Viollier and Trautmann (\cite{vt}), we assume that the
reaction channel {\it (a)} produces photons at energy $E_{\gamma}=
m_{\nu}c^2$ giving rise to an emission line with luminosity
\begin{equation}
\begin{array}{l}
L_{\gamma}^a \simeq 4 \times 10^{25}\left(\frac{M}{2.6\times
10^{6}~M_{\odot}}\right)^3 \times \\
~~~~~~~~~~~~~~~~~~~~~~~~~~~~~~~~~~~~~~~\left(\frac{m_{\nu_{\tau}}c^2}{14.5~
{\rm keV}}\right)^9{\rm erg~s^{-1}}.
\end{array}
\end{equation}
Further, neglecting $X$-ray absorption along the line of sight to
the Galactic Center \footnote{Generally, in order to evaluate the
photon flux on Earth we have also to consider the interstellar
absorption due to the interaction of photons with the
interstellar medium protons. In this way the photon flux on Earth
should be
\begin{equation}
\Phi_{\gamma} \simeq \frac{L_{\gamma}}{4\pi D^2}e^{-\tau}~,
\label{flux}
\end{equation}
where the total optical depth $\tau$ is the sum of the two terms
$\tau_1=\sigma_{\gamma p}n_{1,p}D_1$ and $\tau_2=\sigma_{\gamma
p}n_{2,p}D_2$. The cross section for $\gamma-p$ interaction is
given by (Morrison and McCammon \cite{mm})
\begin{equation}
\sigma_{\gamma p}\simeq 2.3\times 10^{-22}(E_{\gamma}/{\rm
keV})^{-8/3}$ cm$^{2}
\end{equation}
and the proton number densities $n_{1,p}$ and $n_{2,p}$ are
assumed to be $n_{1,p}\simeq 100$ cm$^{-3}$ within $D_1\simeq 100$
pc from SgrA$^*$ (see also Zane, Turolla and Treves \cite{ztt})
and $n_{2,p}\simeq 1$ cm$^{-3}$ from the central region to Earth
at the distance $D_2\simeq 8.5$ kpc, respectively. As we will
see, in the case of interest ($E_{\gamma}\geq 10$ keV)  the
absorption turns out to be negligible.}, the corresponding
photons flux on Earth turns out to be
\begin{equation}\begin{array}{l}
\nonumber
\Phi_{\gamma}^a\simeq4.5\times10^{-13}\left(\frac{M}{2.6\times
10^{6}~M_{\odot}}\right)^3 \times \\
~~~~~~~~~~~~~~~~~~~~~~~~~~~~~~~~~~~~~~~~~\left(\frac{m_{\nu_{\tau}}c^2}{14.5
{\rm keV}}\right)^8{\rm cm^{-2}s^{-1}}.
\end{array}
\end{equation}
Another possible neutrino ball signature comes from the {\it (b)}
reaction channel in (\ref{channels}). In fact, as a consequence of
the standard electroweak interaction theory, heavy neutrinos may
decay into lighter neutrino species trough the emission of a
photon producing an $X$-ray emission line at energy
$E_{\gamma}=m_{\nu_{\tau}}c^2/2$. The corresponding photon
luminosity is
\begin{equation}
L_{\gamma}^b\simeq \frac{Mc^2}{2T_D}~, \label{lumgamma}
\end{equation}
where the radiative decay constant $T_D$ is given by (Viollier,
Leimgruber and Trautmann \cite{vlt})
\begin{equation}
T_D^{-1} \le 3 \times 10^{-19}~
\left(\frac{m_{\nu_{\tau}}c^2}{14.5~{\rm keV}}\right)^5~~{\rm
yr^{-1}}~. \label{decayconst}
\end{equation}
Consequently, the luminosity entails
\begin{equation}
\begin{array}{l}
\nonumber L_{\gamma}^b \leq 2.3\times 10^{34}
\left(\frac{M}{2.6\times 10^{6}~M_{\odot}}\right)\times \\
~~~~~~~~~~~~~~~~~~~~~~~~~~~~~~~~~~~~~~~\left(\frac{m_{\nu_{\tau}}c^2}{14.5
~{\rm keV}}\right)^5~{\rm erg~s^{-1}}~,
\end{array}
\end{equation}
corresponding to a photon flux on Earth
\begin{equation}
\begin{array}{l}
\nonumber \Phi_{\gamma}^b\leq1.2\times
10^{-4}\left(\frac{M}{2.6\times 10^{6}~M_{\odot}}\right)\times \\
~~~~~~~~~~~~~~~~~~~~~~~~~~~~~~~~~~~~~~~\left(\frac{m_{\nu_{\tau}}c^2}{14.5
{\rm keV}}\right)^4~{\rm cm^{-2}s^{-1}}. \label{fluxx}
\end{array}
\end{equation}

Let us now compare this expected photon flux with the presently
available data. In the energy range of interest (10 keV
$\ut<E_{\gamma}<$ 787 keV), data have been collected mainly by the
instruments ART-P on board of GRANAT and OSSE on CGRO.

ART-P made detailed observations towards the Galactic Center
region in the energy band $3$ keV - $30$ keV (Pavlinsky, Grebenev
and Sunyaev \cite{artep}) and in particular towards SgrA$^*$ with
exposure time $\simeq$ 164,000 s. The derived photon spectrum is
well described by a power law model with index $\alpha=-1.6 \pm
0.1$ and the measured average flux, in the $3-20$ keV energy band,
is $8.8\times 10^{-3}$ photons cm$^{-2}$ s$^{-1}$. The absence of
lines and/or of other particular features is clear from Figure
$6$ on Pavlinsky, Grebenev and Sunyaev (\cite{artep}).

OSSE instrument on the CGRO satellite has also observed the
Galactic Center in the energy range $30$ keV - $1$ MeV  with
exposure time of about one day (Smith, Leventhal, Gehrels et al.
\cite{osse}). The observed SgrA$^*$ spectrum can be well fitted
by a power law model with index $\alpha=-2.1 \pm 0.1$ and with
average flux, in the $30-600$ keV energy band, of $9.7\times
10^{-5}$ photons cm$^{-2}$ s$^{-1}$ (Smith, Leventhal, Gehrels et
al. \cite{osse}). It is important to note that in the OSSE
spectrum of the Galactic Center only two structures were
observed, {\it i.e} the emission line at 511 keV, corresponding to
the electron-positron annihilation radiation, and the emission
feature at 170 keV which is interpreted as the Compton
backscattered $511$ keV radiation (Smith, Leventhal, Gehrels et
al. \cite{osse}).

A flux $\Phi_{\gamma}$, due to an emission line at energy
$E_{\gamma}$, is detectable with $k\sigma$ statistical detection
threshold if
\begin{equation}
\Phi_{\gamma}>
k~\sqrt{\frac{\Phi_{\gamma}^{obs}(E_{\gamma})}{T_{obs}}}~,
\label{obs}
\end{equation}
where $T_{obs}$ is the observation time and
\begin{equation}
\Phi_{\gamma}^{obs}(E_{\gamma}) = \int_{E_{\gamma}-\Delta
E}^{E_{\gamma}+\Delta E} \frac{dN}{dE}~dE
\end{equation}
is the observed flux on Earth. Here $dN/dE$ is the photon
spectrum as observed by the ART-P and OSSE satellites and
$2\Delta E$ is the instrument spectral resolution (see Pavlinsky,
Grebenev and Sunyaev \cite{artep}, Molkov, Grebenev, Pavlinsky
and Sunyaev \cite{artep2}, Johnson, Kinzer, Kurfess and Strickman
\cite{jkks}).

\begin{figure}[htbp]
\begin{center}
\vspace{7.8cm} \includegraphics{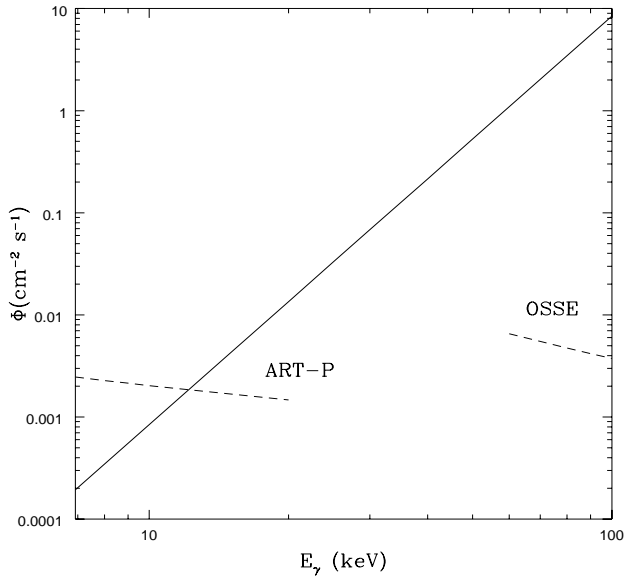} \caption{The photon flux on Earth, produced
by the decay of massive neutrinos into lighter species throughout
a photon emission, is reported for different constituting
neutrino mass. The neutrino ball mass has been assumed to be
$M=(2.6\pm0.2)\times 10^6$ M$_{\odot}$. The dashed lines show the
threshold flux necessary to see a line at energy $E_{\gamma}$
within 5$\sigma$ statistical detection confidence level for the
instruments ART-P and OSSE. The continuous line gives the
expected photon flux in a line at energy
$E_{\gamma}=m_{\nu}c^2/2$ (for details see text).
 }
\label{fluxgamma}
\end{center}
\end{figure}

Clearly, due to the neutrino mass range 11 keV $\ut< m_{\nu}\ut<$
787 keV (obtained from the analysis in Figure \ref{rcontrom}), the
photon flux $\Phi_{\gamma}^b$ in equation (\ref{fluxx}) is the
most favorite signature for direct observation. Thus, by setting
$\Phi_{\gamma}(E_{\gamma})\equiv\Phi_{\gamma}^b(m_{\nu}c^2/2)$,
the two hand sides of equation (\ref{obs}) are plotted in Figure
(\ref{fluxgamma}) as a function of the line energy $E_{\gamma}$.
Inspection of this figure allows us to reject neutrino ball
models corresponding to neutrinos with mass $m_{\nu}\ut >$ 24 kev
since they are expected to imply emission lines which should have
been detected by ART-P. Consequently, the permitted neutrino mass
range reduces to 11 keV $\ut < m_{\nu}\ut <$ 24 keV.
%
\section{Discussion and conclusion}

In this paper, we investigate the possibility that the Galactic
Center hosts a massive neutrino ball of total mass $M\simeq
2.6\times 10^6$ M$_{\odot}$. The existence of such an object,
under particular circumstances, should justify the low
luminosity, from radio to $\gamma$-rays, observed in the
direction of SgrA$^*$.

We build up models for the neutrino ball by studying the
gravitational equilibrium of a semi-degenerate fermion gas.
Density and pressure within the ball are defined by adopting a
formalism based on a distribution function in phase space, which
allows us to consider neutrinos with a degeneracy degree varying
from the center to the border of the system. Limiting cases are
the fully degenerate fermion systems (which are represented by
the oblique dashed line in Figure \ref{rcontrom}) and the
classical isothermal spheres well known in the literature. The
local balance between gravitational force and pressure gradient
leads to stable configurations if the number of neutrinos (and/or
anti-neutrinos) does not exceed the critical value in equation
(\ref{maxnumber}). This fact, for a total ball mass $M \simeq 2.6
\times 10^6~M_{\odot}$, allow us to put an upper limit to the
neutrino mass $m_{\nu}\ut< 787$ keV. This limit is represented in
Figure \ref{rcontrom} by the vertical solid line. Acceptable
neutrino ball models in Figure \ref{rcontrom} are those between
the oblique dashed line and the vertical solid one, having
decreasing degeneracy with increasing neutrino mass $m_{\nu}$.

By requiring, moreover, that the observed luminosity towards
SgrA$^*$ comes from the accretion process on the neutrino ball and
that the evaporation time scale of the ball is longer than the
Hubble time, the allowed models are those on the grey region in
Figure \ref{rcontrom}. Correspondingly, we get that 11 keV $\ut<
m_{\nu}\ut< 787$ keV.

The above neutrino mass range can be further reduced by studying
the photon flux on Earth due to {\it i)} interaction of incoming
ultra high energy neutrinos (or anti-neutrinos) with
anti-neutrinos (or neutrinos) composing the ball, and {\it ii)}
interaction between neutrinos and anti-neutrinos in the ball
itself (\ref{channels} a) and $\tau$ neutrino decay
(\ref{channels} b). Investigation of such effects gives us the
opportunity to test the model itself by comparing the neutrino
ball signature with the available satellite observations. In
particular, the neutrino decay reaction $\nu_{\tau} \rightarrow
\nu_{e}+\gamma$ gives rise to an emission line at energy
$m_{\nu_{\tau}}c^2/2$. However, a detailed analysis of the
observed spectrum towards the Galactic Center allows us to
exclude such a signal for a constituting neutrino mass $m_{\nu}
\ut> 24$ keV (see Figure \ref{fluxgamma}). Therefore, present
observations do not allow to exclude the existence of a neutrino
ball at the Galactic Center with mass $M\simeq 2.6\times 10^6$
$M_{\odot}$ if the constituting neutrino mass $m_{\nu}$ is in the
range 11 keV $\ut<m_{\nu}\ut<$ 24 keV. The next generation of
$X$-ray satellites, like XEUS (XEUS home-page \cite{xeus}) and
Constellation-X (Constellation-X home-page \cite{constx}), with
improved sensitivity and angular resolution will be able to
definitively exclude or confirm the existence of a neutrino ball
with constituting particle mass in the above range.
\acknowledgements{We thank Dr. Daniele Montanino for interesting
discussions.}

\end{document}